# Fetal ECG Extraction from Maternal ECG using Attention-based CycleGAN

Mohammad Reza Mohebbian*, Seyed Shahim Vedaei, Khan A. Wahid, *Member, IEEE*, Anh Dinh, *Member, IEEE*, Hamid Reza Marateb, *Member, IEEE*, Kouhyar Tavakolian, Senior *Member, IEEE*

*Abstract*— Non-invasive fetal electrocardiogram (FECG) is used to monitor the electrical pulse of the fetal heart. Decomposing the FECG signal from maternal ECG (MECG) is a blind source separation problem, which is hard due to the low amplitude of FECG, the overlap of R waves, and the potential exposure to noise from different sources. Traditional decomposition techniques, such as adaptive filters, require tuning, alignment, or pre-configuration, such as modeling the noise or desired signal. to map MECG to FECG efficiently. The high correlation between maternal and fetal ECG parts decreases the performance of convolution layers. Therefore, the masking region of interest using the attention mechanism is performed for improving signal generators' precision. The sine activation function is also used since it could retain more details when converting two signal domains. Three available datasets from the Physionet, including A&D FECG, NI-FECG, and NI-FECG challenge, and one synthetic dataset using FECGSYN toolbox, are used to evaluate the performance. The proposed method could map abdominal MECG to scalp FECG with an average 98% R-Square [CI 95%: 97%, 99%] as the goodness of fit on A&D FECG dataset. Moreover, it achieved 99.7 % F1-score [CI 95%: 97.8-99.9], 99.6% F1-score [CI 95%: 98.2%, 99.9%] and 99.3% F1-score [CI 95%: 95.3%, 99.9%] for fetal QRS detection on, A&D FECG, NI-FECG and NI-FECG challenge datasets, respectively. These results are comparable to the state-of-the-art; thus, the proposed algorithm has the potential of being used for high-performance signal-to-signal conversion.

## I. INTRODUCTION

The electrocardiogram (ECG) signal is used as a non-invasive method of heart disorders diagnosis. Connecting electrodes on the chest, arms, hands, or legs is the traditional way of recording ECG [1]. In addition, ECG is used for fetal heart monitoring [2].

There are two ways of recording a fetal electrocardiogram (FECG). In invasive recording FECG, electrodes are attached to the fetal scalp during delivery. In this way, the risk of infection exists; however, the signal quality is excellent. In non-invasive FECG recording, maternal abdominal ECG can be utilized for FECG extraction. Emerging these extraction techniques have revolutionized fetus healthcare and enabled the clinician to monitor fetal heart activities continuously. A proper fetal heart rate involves normal mother oxygenation and transferring oxygen to the fetus. Every mechanism that induces an oxygen channel breakage may cause disturbances in the fetal heart rate.

The obtained MECG signal may be distorted with noise (baseline drift and motion artifacts) [2]. These noise sources may contribute to measurement and instrumentation failures, such as power-line disturbance, white noise, electrode connection noise, muscle contraction, electrosurgical noise, baseline wandering, and motion anomalies [3]. Slight distortion in the MECG waveform can impair the FECG extraction [4]. Therefore, using a robust decomposition strategy against noise helps extract and represent the fetus's heart functionality. The majority of the decomposition approaches attempt to extract FECG's QRS waves [5], [6]; however, FECG has other essential components, which can allow extensive analysis [7]. The most recognizable parts in FECG are P waves (the depolarization wave from the SA node that traverses the atria), QRS complex (ventricle depolarization), ST segments (both ventricles are depolarized completely), and T wave (ventricular repolarization) [8].

Adaptive filters are essential methods for extracting FECG components, wherein coefficients are adapted according to the signal changes in time. However, the power spectral density of the input signal affects the convergence rate of the algorithms [9]. Since the minimum mean-square error is mostly used in adaptive filter's objective function, they require a flat and uniform power spectrum to have excellent convergence. However, real-world problems include colored noise rather than white noise [10], and colored noise can drop the efficiency of adaptive filters. The least mean squares (LMS) and Recursive least square (RLS), adapted like Weiner optimal solution, are designed to work with narrowband frequencies [9]. The prerequisite of RLS and LMS is that they require a reference signal that is morphologically akin to the abdominal

Mohammad Reza Mohebbian, Seyed Shahim Vedaei, Khan A. Wahid and Anh Dinh are with Department of Electrical and Computer Engineering, University of Saskatchewan S7N 5A9, Saskatoon, Saskatchewan, Canada.

Hamid Reza Marateb is with Biomedical Engineering Department, University of Isfahan, Isfahan, Iran, and also with Department of Automatic Control, Biomedical Engineering Research Center, Universitat Politècnica de Catalunya, BarcelonaTech (UPC), Barcelona, Spain.

Kouhyar Tavakolian is with the School of Electrical Engineering and Computer Science, University of North Dakota, Grand Forks, ND 58202, USA.

Corresponding author: Mohammad Reza Mohebbian (e-mail: mom158@usask.ca).

The authors declare that the research was conducted in the absence of any commercial or financial relationships that could be construed as a potential conflict of interest.

MECG waveform. Methods relying on temporal features, like template-based and conventional Kalman filters, are other approaches that may be listed as failing when MECG and FECG peaks overlap. The extended state Kalman filter was introduced for robust FECG extraction, which could solve the QRS coincidence issue [11]. Nevertheless, they have computational complexity and unable to succeed if they could not accurately detect R-peaks.

Blind source separation strategies such as principal component analysis (PCA), independent component analysis (ICA), and periodic component analysis are substitutes of adaptive filters [12]. The primary concept of these approaches is a linear stationary mixing matrix between sources and the higher number of abdominal channels for better FECG extraction [13]. However, these methods are not adequate in insufficient Signal to Noise Ratio (SNR) circumstances and usually require specific electrode configuration and further post-processing [14].

Some researchers attempted to overcome the adaptive filter or blind source separation (BSS) drawbacks using novel techniques. Mohebbian *et al.* [14] used a BSS technique to estimate the reference signal on the adaptive filter and decompose FECG using one channel with F1-Score 96%. Zhang *et al.* [15] used singular value decomposition and smooth window and could decompose QRS of FECG with F1-Score 99%. A convolutional neural network was utilized by Zhong *et al.* [5] and could achieve 77% F1-Score for QRS wave extraction. QRS detection was also performed by Zhong *et al.* [16] using the prefix tree-based model and could achieve the F1-Score of 95%. Many methods tried to extract QRS waves; however, in decomposing FECG, other parts should also be taken into consideration. Moreover, using a general model that can be used for different subjects with different electrode displacement needs more investigation.

Simulating MECG tries to cover all artifacts in synthetic signals to enable algorithms to be tested on different scenarios, such as various noises, gestational ages, and assorted artifacts [17]. One of the famous libraries for generating synthetic MECG is fecgsyn [18], which uses realistic noise, a heart rate variability probability model, rotation maternal and fetal heart axes, fetal movement, and physiological features. The generation of FECG signals based on physiological and mathematical models needs a systematic understanding of the factors involved in producing a FECG signal. However, evolving generative adversarial approaches recently provided new perspectives for the latent vector-based data generation that learns the system's nature without prior information [19].

In this paper, the attention-based Cycle Generative Adversarial Network (CycleGAN) is introduced for mapping between MECG and FECG. The idea of CycleGAN is from Hertzmann *et al.* [20], who utilized a non-parametric texture model for mapping two images. Using the weight-sharing technique for representing across domains was used by Aytar *et al.*[21] and was extended by Liu *et al.*[22] by adopting variational autoencoders and generative adversarial networks. Recently, CycleGAN has been used for non-parallel voice conversion [23], adapting image emotion [24], and style transfer in x-ray angiography [25]. Also, using deep neural networks for ECG analysis has been investigated in recent studies [26]–[28]. We modified CycleGAN to use attention and sine activation to find a map between maternal and fetal ECG using adversarial loss. The proposed algorithm is trained and evaluated using three datasets provided by Physionet. The remaining of the paper is arranged as follows: information about the datasets and method implementation are presented in the next section. The results of the proposed approach are described in section 3. Sections 4 and 5 include the discussion and conclusions, respectively.

## II. MATERIALS AND METHOD

Figure 1 shows the proposed method in detail. Briefly, a sliding window is applied to the abdominal MECG and corresponding FECG signals. Then, they are normalized, and bandpass filtered and feed to the CycleGAN network. Two generators are trained to receive two discriminators on an adversarial training strategy. The generators are designed based on one-dimensional signal processing and should maintain the smoothness of the signal. The proposed method is evaluated in two different scenarios, including FECG signal extraction and fetal QRS detection. In FECG signal extraction, whole FECG signal components are extracted, and signal distortion is essential, while in fetal QRS detection, only R-wave positions are extracted, and accuracy of R-R detection is crucial. More specific details regarding each phase are given in the sub-sections below.

### 2.1 Dataset
#### 2.1.1 A&D FECG

The abdominal and direct FECG (A&D FECG) [29] dataset from Physionet was used as the main dataset that contains FECG recordings. The data contains multichannel fetal electrocardiogram (FECG) recordings obtained from a fetus scalp of 5 different women (subject 1: record r01, subject 2: record r07, subject 3: record r10, subject 4: record r04, and subject 5: record r08), between 38 and 41 weeks of gestation. Each recording consists of five minutes, four abdominal channels, and a corresponding FECG obtained from the fetal head. All signals were sampled at 1 kHz and 16-bit resolution and bandpass filtered during acquisition (0-100 Hz) with digital filtering of the power-line. The abdominal electrode configuration consisted of four electrodes around the belly button, a reference electrode above the symphysis of the pubic, and a common reference electrode on the left leg. These positions were constant during all recordings. Although the FECG is recorded directly from the scalp, it contaminates the maternal ECG. According to Nurani *et al.* [30], when the period is closer to delivery due to uterine contractions, the impact of maternal ECG on directly recorded FECG can also be higher.

#### 2.1.2 NI-FECG and NI-FECG challenge

The non-invasive FECG (NI-FECG) Physionet was also used [31]. This dataset does not contain direct FECG signals from the scalp and only have QRS time samples for FECG. It contains 55 multichannel abdominal MECG, taken from a subject between 21 and 40 weeks of pregnancy. The electrode positions are not fixed and change sometimes to change the signal to noise ratio. The data is sampled at 1 kHz and 16-bit resolution and bandpass filtered during acquisition (0-100 Hz). Fourteen sets, including 154, 192, 244, 274, 290, 323, 368, 444, 597, 733, 746, 811, 826, 906, are selected because other research selected almost the same sets and it facilitates the comparison [12], [14].

Set-A of the 2013 Physionet/Computing in Cardiology Challenge [32] is used for benchmarking [32]. The archive consists of 75 abdominal ECG data recorded at a sampling rate of 1 kHz on four channels. The fetal reference R-peaks are provided for Set-A. According to many incomplete annotations [33], [34], records a33, a38, a52, a54, a71, and a74 were excluded, leaving 69 records for evaluation.

In order to train the proposed method on NI-FECG and NI-FECG challenge datasets, simulated FECG is generated based on R-R interval provided as ground truth in datasets. More details about simulation which is based on Daubechies wavelets is provided in [35]. These simulated signals are mainly used for the training of the proposed system to be used in QRS detection and no further assessment is carried out for the quality of FECG extraction such as what is performed on the A&D FECG dataset.

*2.1.3 Simulated Signals:*

The FECGSYN toolbox [36] is used, in addition to the available datasets, to model 24 maternal and fetal ECGs at various fetal and maternal heart rates, including 115, 125, 135, 145, 155, 160 bpm [37] and 65, 77, 89, 100 bpm, respectively [38].

*2.2 Preprocessing*

Since the first and last parts of each record have more artifacts and noise, 10 seconds from the first and the last parts are truncated. Signals are resampled to 200 Hz to have less computational cost and the same sample numbers in each second. The clinical ECG has a minimum bandwidth of 100 Hz [39], and according to Sameni and Clifford [40], the majority of the ECG relative power falls under 35 Hz, and the QRS complex has a frequency range of 10-15 Hz for FECG. Therefore, a bandpass filter with the cut-off frequencies 1 Hz and 100 Hz is applied to signals. Since the smoother signal offers better results for the proposed approach and it is not acceptable to select a narrower bandpass filter owing to losing information, the Savitzky-Golay filter [41], [42] was used in our study. The rectangular sliding window is performed on all signals; then, signals are normalized using z-score normalization [43]. The vector $x_i$ is referred to the one abdominal MECG channel in size 1×M. Also, *i* represents the *i-th* window. Similarly, $y_i$ represents FECG vector, for *i-th* window with size the 1×M, wherein M refers to the number of samples, which is set to 200 for this paper.

*2.3 The Proposed Method*

The goal is to map domains $X = \{x_1, x_2, ..., x_N\}$ and $Y = \{y_1, y_2, ..., y_P\}$, while $X$ is $N \times M$ abdominal MECG and $Y$ is the $P \times M$, containing the FECG signal. Like other CycleGAN approaches [23], the model contains two mappings, including $G: X \rightarrow Y$ and $F: Y \rightarrow X$. Also, two adversarial discriminators $D_x$ and $D_y$ should be defined, where $D_x$ is focused to differentiate between $X$ and $F(Y)$ and objective of $D_y$ is to discriminate between $Y$ and $G(X)$. Concisely, two main objectives follow adversarial loss, trying to fit the representation of the produced signal to represent data in the endpoint domain, and cycle consistency loss that avoids the trained mappings G and F from contradicting each other. The final objective cost is defined in Equation (1).

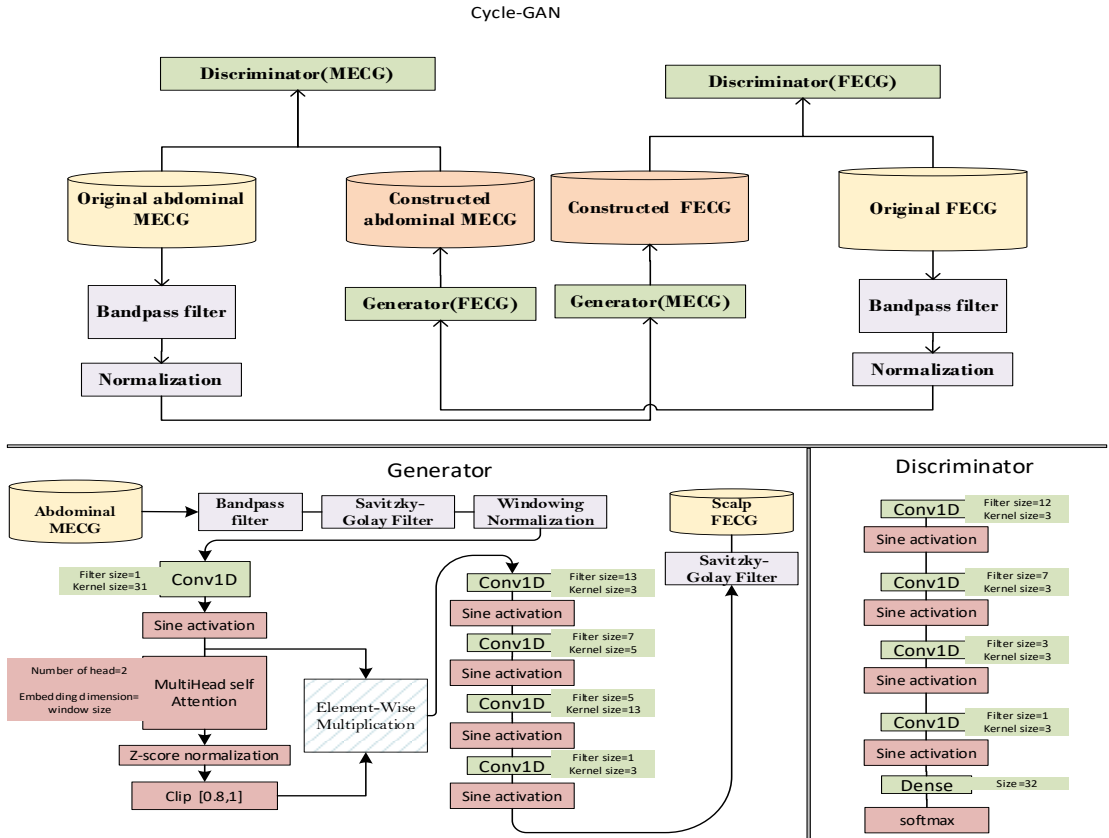

Figure 1. The high-level block diagram of the proposed algorithm (top); The generator used in CycleGAN (bottom left); the discriminator used for CycleGAN (bottom right).

$$\arg\min_{G,F}\max_{D_x,D_y} \mathcal{L}(G,F,D_x,D_y)$$
$$\mathcal{L}(G,F,D_x,D_y) = \mathcal{L}_{GAN}(G,D_y,X,Y) + \mathcal{L}_{GAN}(F,D_x,Y,X) + \lambda \mathcal{L}_{cyc}(G,F) \quad (1)$$

Wherein,
$$\mathcal{L}_{GAN}(G,D_y,X,Y) = E_y[\log D_y(y)] + E_x[\log(1 - D_y(G(x)))] \quad (2)$$

$$\mathcal{L}_{GAN}(F,D_x,Y,X) = E_x[\log D_x(x)] + E_y[\log(1 - D_x(F(y)))] \quad (3)$$

$$\mathcal{L}_{cyc}(G,F) = E_x[\|F(G(x)) - x\|_1] + E_y[\|G(F(y)) - y\|_1] \quad (4)$$

where $\lambda$ controls relative objectives, which is set to 4 to give more weight to cyclic consistency. Equations 2 and 3 represent the adversarial objective, which trains $G$ and $F$ to produce outputs identically as targets. On the other hand, objective 4 has two terms of forward and backward cycle consistency that guarantees the learned models can map individual input to a specific output. Using greater lambda reduces identity mapping.

Former GAN methods depend on the presence of one-to-one examples for training; however, the CycleGAN can learn these transformations without the need to map one-to-one between the training data in source and target domains [23]. A coupled signal requirement in the target domain is removed by having a two-step transition-first by attempting to convert it to the target domain and then back to the main signal. The generator is applied to the signal to map it to the target domain, then the quality of the generated signal is improved by performing the generator against a discriminator. This feature enables the algorithm to train with a small dataset.

The network architecture contains two main parts, generator and discriminator. Both generator and discriminator architectures are depicted in Figure 1. The purpose of the generator is to intensify waves related to the target signal and to reduce the influence of waves of input signals. However, based on the analysis, we observed that the convolutional kernels, which could increase the amplitude of the FECG waves, could also boost the MECG waves. In other words, instead of suppressing the maternal R wave, it may amplify it along with enhancing the R wave of the fetus. To solve this problem, an attention layer [44] is used to provide a mask to certain parts of the signal and avoid processing parts that can increase errors. Instead of using the softmax layer, which is usual after attention, the output of the attention layer is normalized, and weights that are between 0.8 and 1 are selected for masking. The mask is multiplied to the signal to extract the region of interest. Next, three one-dimensional convolution layer (Conv1D) are applied with the sine activation function [45]. The sine activation function has shown significant results for representing the signals such as audio, video, and images and can retain details better than other popular activation functions like LeakyRelu. The three Conv1D aim to enhance FECG and finally smooth signals using stride two and bigger kernel size parameters [46]. The discriminator contains four Conv1D layers with a sine activation function that tries to classify the input as fake or real using a dense layer and softmax as classification. The generator is trained to produce signals that the discriminator identifies as actual. The algorithm ran for 50 epochs with the loss function $\log(\cosh)$ and Nesterov Adam optimizer [47].

### 2.4 Validation

The performance of the proposed method is evaluated in two ways: quality of extracted signal and accuracy of QRS detection. In all cases, following the STARD [48] and TRIPOD [49] standards, the CI 95% of the performance indices are reported for showing the reliability of the estimation.

#### 2.4.1 Signal extraction quality

In the A&D FECG dataset, the FECG signals are recorded from the fetus's scalp and can be used as continuous ground truth. For NI-FECG and NI-FECG challenge datasets, there is no recorded FECG, and the simulated FECG based on R information is only used for QRS evaluation. The leave-one-subject-out cross-validation approach was used for performance assessment. The entire signals except for one subject are used for the train, and the remaining signal is used for tests. This approach is repeated five times to test all subjects. The R-squared goodness of fit [50], intra-class correlation (ICC) [51], and Bland-Altman plots or Tukey's Mean Difference Map [52] are used for reporting the performance on each subject. The Tukey's Mean Difference Map is a statistical tool for evaluating the variations between the two measurement processes.

For diagnostic distortion analysis purposes, the mean value of the original signal and the predicted signal are subtracted from the signals. Using Daubechies 9/7 biorthogonal Wavelet filters up to 5 levels, all signals are decomposed. The QRS complex typically has the highest amplitude and the broadest spectrum. Therefore, the QRS complex is visible on all levels, however it is more noticeable in the second and third levels. On the first two levels, the P and T waves are not apparent and do primarily belong to four and five levels. The deviation between the original signal's Wavelet coefficients and the reconstructed signal's Wavelet coefficients is determined by the percentage root mean square difference, referred to as Wavelet PRD (WPRD). Finally, the Wavelet Energy-based Diagnostic Distortion (WEDD) is calculated by the weighted average of WPRD in all levels [53] as shown in equation 5.

$$WEDD = \sum_{l=1}^{L+1} w_l \ WPRD_l = \sum_{l=1}^{L+1} w_l \sqrt{\frac{\sum_{k=1}^{K_l}[d_l(k) - \widetilde{d_l(k)}]^2}{\sum_{k=1}^{K_l}[d_l(k)]^2}},$$
$$l = 1,2,3, \ldots L \quad (5)$$

where, $WPRD_l$ is the error in $l-th$ subband, and $d_l(k)$ and $\widetilde{d_l(k)}$ are the $k-th$ wavelet coefficient in $l-th$ subband of original and predicted signals, respectively. The WEDD value can be categorized into five quality groups [53], including excellent (0-4.6), very good (4.6-7), good (7-11.2), not bad (11.2-13.6), and bad (>13.6).

The paired-sample t-test was also used to identify whether the reconstructed signals are significantly different from the original signals, especially in terms of bias. Results are reported as mean ± standard deviation, and P-values less than 0.05 were considered significant.

*2.4.2 QRS detection accuracy*

The QRS wave detection was performed using P&T [54] and analyzed using traditional sensitivity, positive predictive value (PPV), and F1-Score [2], [15], [16]. For calculating the performance, different time precision is used by various researchers. For example, Guerrero-Martinez *et al.* [55] used a matching window of 50 ms, while Zhang *et al*. [15] used 30 ms. In this study, every one sample in decomposed data belongs to the 5 ms due to down-sampling. Therefore, a window of 6 samples (30 ms) is used for calculating performance indices.

The statistical analysis was performed using SPSS Statistics for Windows version 22 (IBM Corp. Released 2013. Armonk, NY: IBM Corp.). All algorithms are run on a system with Core-i9, 16 GB RAM, and 6 GB Graphic Cards NVIDIA GeForce GTX 1060 configurations.

## III. RESULTS

*3.1 Results for signal extraction*

Two examples of mapping MECG and FECG is depicted in Figure 2 for subject 2 and 4.

The ICC, R- squared and WEDD indices are shown in Table I. All parameters suggest that the signals predicted are very similar to the original ones. Moreover, the Tukey/Bland-Altman mean difference plot of the decomposed FECG for each subject is depicted in Figure 3.

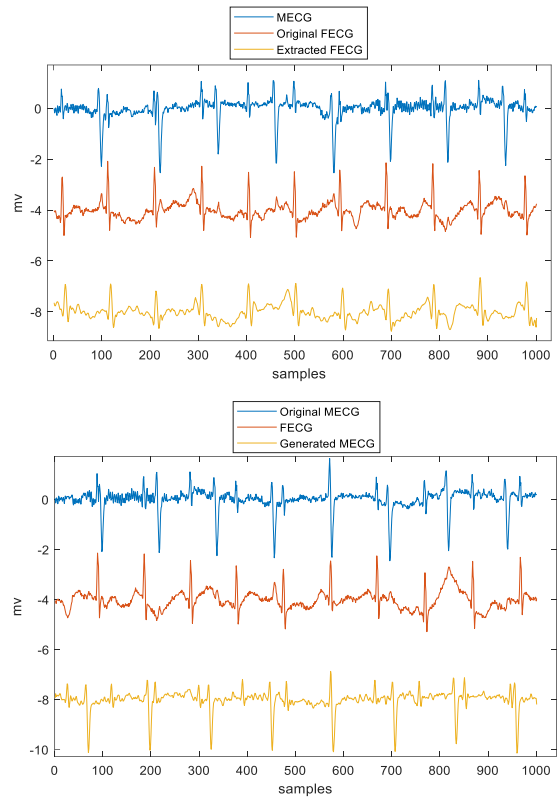

Figure 2. Two examples of FECG and MECG generation. The top one is the decomposed FECG using proposed method on 1000 samples of MECG test set for subject 4; The bottom one is the generated MECG from FECG on subject 2 for 1000 samples. All signals are normalized for visualization purposes.

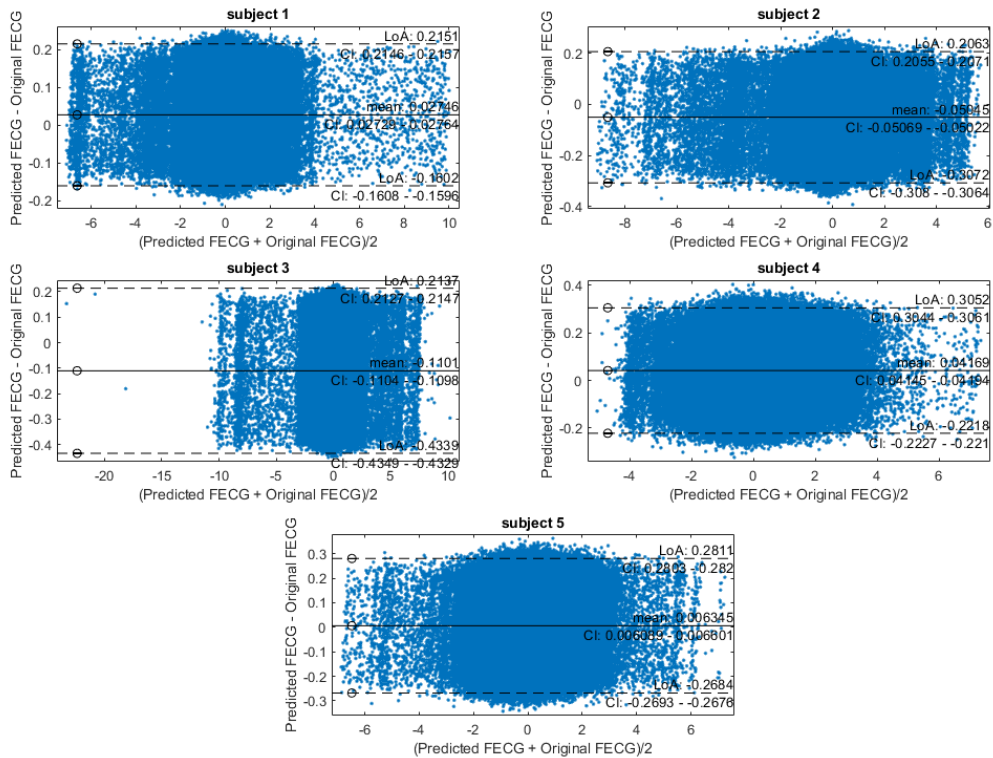

Figure 3. The mean difference plots between predicted FECG and original FECG for all subjects in A&D FECG dataset.

TABLE I
THE PERFORMANCE OF THE ESTIMATED FECG SIGNAL (CI 95% ARE REPORTED IN PARENTHESIS)

| Subject | 1 | 2 | 3 | 4 | 5 |
|---|---|---|---|---|---|
| ICC | 0.99 (0.97-0.99) | 0.99 (0.97-0.99) | 0.98 (0.94-0.99) | 0.99 (0.95-0.99) | 0.99 (0.97-0.99) |
| R-squared | 0.99 (0.99-1.00) | 0.98 (0.97-0.99) | 0.96 (0.93-0.97) | 0.98 (0.97-0.99) | 0.98 (0.97-0.99) |
| WEDD | 7.0 % (6.3–7.3) | 5.7 % (4.7-5.8) | 9.2% (8.1-9.9) | 6.3% (6.1-6.8) | 7.2% (6.8-8.0) |

According to [53] and WEDD values, the rate of distortion is in "very good" and "good" ranges for all subjects, which means that the main FECG components are retained for all subjects. Moreover, there were no significant differences between the reconstructed and original signals 1-5 (paired t-test; P-value>0.1).

Since the training and assessment of A&D FECG dataset was based on subject-leave-out, not all heart rates were considered for the trained models. To generalize, a model is trained on all the A&D FECG dataset. This model is used for testing on NI-FECG, and NI-FECG challenge datasets. Moreover, this model is applied on 24 simulated MECG and FECG signals generated by the FECGSYN toolbox. The boxplot for R-Squared and WEDD indices are plotted for different maternal and fetal heart rates in Figure 4. This figure shows that the system's performance does not substantially drop if it is trained with data that cover most of the heart rate variabilities. There is no correlation between WEDD, and maternal HR (Kendall's $\tau_b$ =-0.016; P-value=0.918). However, there is a weakly negative correlation between R-square and maternal HR but not statistically significant (Kendall's $\tau_b$ =-0.221; P-value=0.165).

### 3.1 Results for QRS detection

NI-FECG challenge and NI-FECG are used for QRS detection evaluation. NI-FECG is obtained from a subject in 21 and 40 weeks of pregnancy. Compared to the A&D FECG dataset recorded from women in 38 and 41 weeks, the FECG QRS of these two datasets has less amplitude, and it was more difficult to extract FECG. In one way, the trained system on the A&D FECG dataset is tested on 14 signals of NI-FECG and 69 signals of NI-FECG challenge. In the second way, 4-fold cross-validation is performed on the NI-FECG dataset and NI-FECG challenge datasets, and the average and standard deviation of test sets are reported. As explained earlier, since the original FECG is not provided in these two datasets, simulated FECG is generated based on Daubechies wavelets, and R-R interval values provided as ground truth are used as the parameter. Table II shows the results.

TABLE II
PERFORMANCE OF THE QRS DETECTION BASED ON P&T METHOD; MEAN ± STD (CI 95% IS REPORTED IN PARENTHESIS)

| Train | Test | F1-score (%) | PPV (%) | Sensitivity (%) | # tested signals |
|---|---|---|---|---|---|
| A&D FECG | A&D FECG | 99.7 ± 0.4 (97.8-99.9) | 99.6 ± 0.7 (97.4 - 99.9) | 99.4 ± 0.6 (98.8- 99.7) | 5 |
|  | NI-FECG | 97.9 (96.5-98.4) | 97.2 (95.9-97.9) | 96.8 (95.4-98.1) | 14 |
|  | NI-FECG challenge | 94.7 (92.6-96.5) | 93.9 (92.6-95.3) | 93.9 (91.6-95.2) | 69 |
| NI-FECG | NI-FECG | 99.6 ± 1.3 (98.2-99.9) | 99.6 ± 0.6 (98.9-99.9) | 99.6± 0.9 (98.8-99.9) | 14 |
| NI-FECG challenge | NI-FECG challenge | 99.3 ± 1.3 (95.3-99.9) | 99.1 ± 1.0 (95.3-99.9) | 99.2 ± 0.6 (96.1-99.9) | 69 |

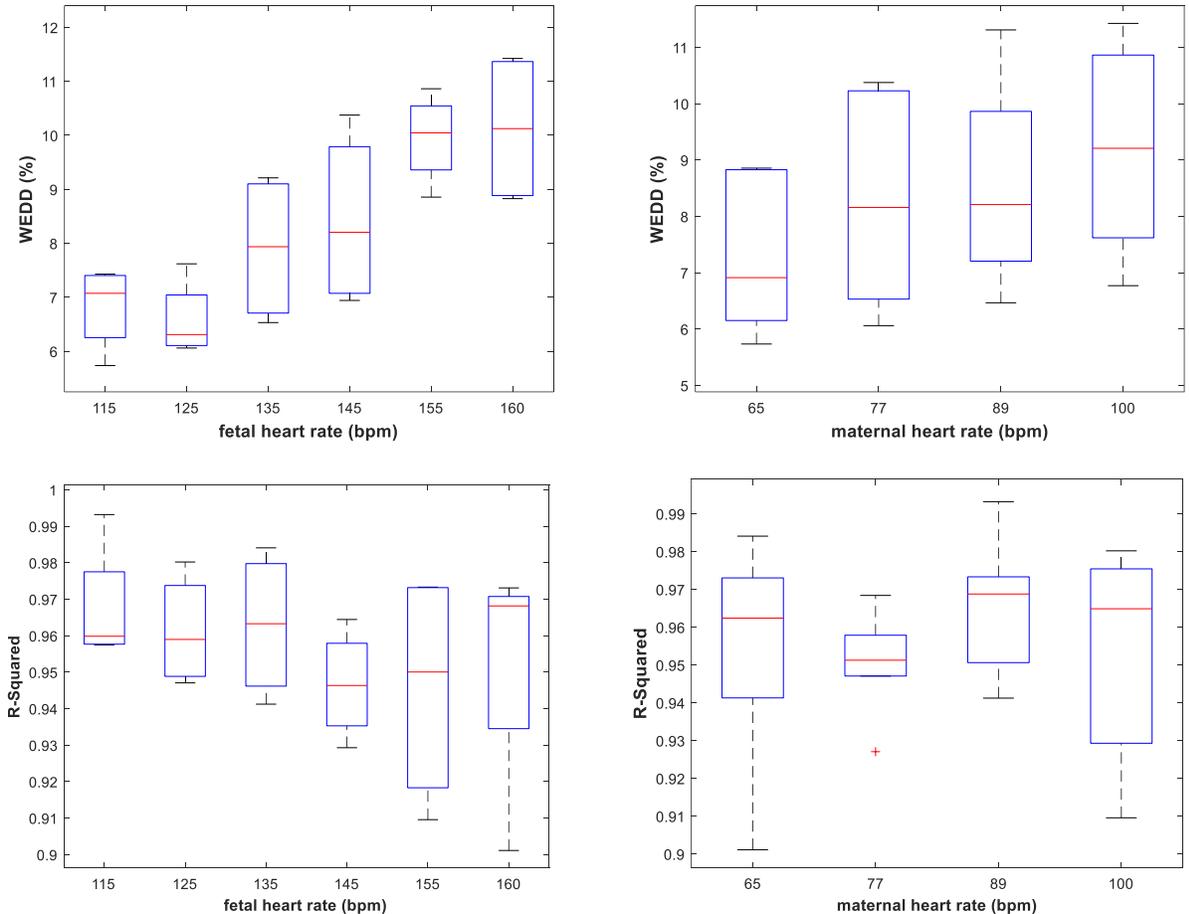

Fig. 4. Boxplots for R-Squared and WEDD indices computed on simulated FECG and MECG based on trained model using A&D FECG dataset. The trained system can extract fetal signal in different heart rates without considerable decreased performance.

## IV. Discussion

A signal-to-signal model is trained for extracting FECG signal from abdominal MECG. The novelty of this algorithm was using the attention mechanism as a filtering mask for focusing on the signal region of interest. Furthermore, using sine activation function and log(cosh) loss was other important parts that have a significant effect on results. To address this argument, each part will be discussed separately.

The proposed method without using the attention layer cannot perform well. A model is trained only based on Conv1D layers along with LeakyRelu activation function and normalization. The structure of the discriminator and all other parameters remained the same. The R-Squared index for five subjects of the A&D FECG dataset is dropped to 0.79, 0.80, 0.77, 0.82, and 0.81, for subjects 1 to 5, respectively. By replacing the Sine activation instead of LeakyRelu the R-Squared increase to 0.80, 0.81, 0.78, 0.83 and 0.81, respectively.

For the loss function, the use of $l_1 = mean\ absolute\ error$, log(cosh) and $l_2 = mean\ square\ error$ is tested. The $l_1$ loss had a better performance than $l_2$ and log(cosh) has better performance than $l_1$. Using $l_2$ norm induces unnecessary smoothing, which causes less efficiency. Since most of the ECG signal contains low-frequency parts, $l_2$ norm may even converge to constant zero values. However, signal fluctuations are preserved by the standard $l_1$. The smoothing effect of $l_2$ was also observed for an electrocardiographic inverse problem for epicardial potential [56]. log(cosh) on the other hand, gives smooth results and preserves signals details. It also has been proved to improve the reconstruction without damaging the latent space optimization [57].

The variation of $\lambda$ coefficient, which is in the objective function of CycleGAN in equation 1, is assessed by adjusting it from 0.1 to 40, and the sensitivity analysis plot is depicted in Figure 5. Concisely, the cyclic and identity of the system could be preserved in 2 < Lambda<10. The higher the Lambda meaning, the less precision in FECG QRS detection. In comparison, using smaller Lambda does not allow the algorithm to converge owing to extra focus on identity.

Other network parameters are also analyzed by utilizing sensitivity analysis. In different steps, kernel size for each layer is changed. The selected kernel size for the proposed method was based on when signal needs more smoothing (bigger kernel size) and when it needs to retrieve more details (smaller kernel size). Sensitivity analysis showed that changing kernel sizes can decrease the performance, however if their ratio respecting to each other remained same, the same performance can be achieved. Finally, the number of Conv1D layers after attention is decreased from 4 to 3 (first Conv1D layer after attention layer is removed), and the average R-Squared performance dropped by 5%.

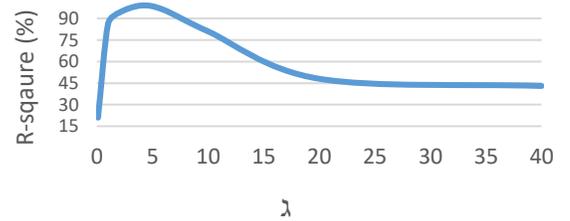

Fig. 5. Sensitivity analysis of Lambda for subject 1 in the A&D FECG dataset.

The proposed method validation was made of two main phases: FECG extraction and QRS detection. Table III compares different approaches for fetal QRS complex detection. It is worth noting external validation shows whether the models are robust enough to work with similar though not identical data. Being in different weeks of gestation for training and test sets results in different quality and different intensity of fetal cardiac activity, which can be stated as the critical factor for dropped accuracy.

The average performance of the proposed method on the NI-FECG challenge dataset is less when it is performed on other datasets. In some noisy abdominal records with small FECG amplitude, like a29, a38, a40, a42, a52, a53, a54, a56, a61, and a65, the performance considerably deteriorated, which affect the overall performance.

Behar et al. [12] evaluated recurrent neural network (RNN), PCA, LMS, and RLS, and template subtraction for QRS detection based on the P&T method [54]. They trained on 30 seconds of the NI-FECG dataset and used a 50 ms matching window. They also reported results of the trained model on a single dataset as external validation. However, the proposed method could achieve better results. The proposed method also gave higher accuracy on the A&D FECG dataset over the Encoder-Decoder method [58] and SVD-SW [15] approaches while tested on more signals. The proposed method outperformed Behar et al. [34] approach that used a combination of template subtraction and ICA. The proposed method also outperformed Varanini et al. [33] QRS detection algorithm. They proposed a signal extraction technique based on ICA and a post-processing method specialized for detecting the QRS on the NI-FECG challenge dataset. Warmerdam et al. [59] used a multichannel hierarchical probabilistic system, which incorporates ECG waveform and heart rate predictive models to detect fetal R peaks in NI-FECG challenge dataset. They reported only 99.6% accuracy defined as TP/(TP+FP+FN) and did not provide sensitivity or F1-Score for better comparison. However, their accuracy outperformed Varanini et al [33] method with an accuracy of 98.6%. The accuracy of our proposed method was 99.6% on QRS detection,

TABLE III
COMPARING THE RESULT OF THE DIFFERENT METHODS FOR DECOMPOSING FETAL ECG

| Method | F1-score (%) | Dataset | QRS detection method | Number of channels | Matching window length (ms) | Number of tested signals |
|---|---|---|---|---|---|---|
| **TS and ES-RNN [12]** | 97.2 | NI-FECG | P&T | 1 Abdominal | 50 | 14 |
| | 90.2 | Train on NI-FECG and test on private dataset | | | | |
| **TS and PCA [12]** | 95.4 | NI-FECG | | | | |
| | 89.3 | Train on NI-FECG and test on private dataset | | | | |
| **LMS [12]** | 95.4 | NI-FECG | | | | |
| | 87.9 | Train on NI-FECG and test on private dataset | | | | |
| **RLS [12]** | 95.9 | NI-FECG | | | | |
| | 88.2 | Train on NI-FECG and test on private dataset | | | | |
| **OBACKC [14]** | 95.6 | NI-FECG | Thresholding | 1 Thoracic | | |
| **SVD-SW [15]** | 99.4 | A&D FECG | P&T | 1 Abdominal | | 2 |
| **Encoder-Decoder [58]** | 94.1 | A&D FECG | | 1 Abdominal | | 5 |
| **Behar et al. [34]** | 95.9 | NI-FECG challenge | | 4 Abdominal | | 69 |
| **Varanini et al. [33]** | 99.0 | NI-FECG challenge | Customized | 4 Abdominal | | 69 |
| **Proposed Method** | 99.6 | NI-FECG | P&T | 1 Abdominal | 30 | 14 |
| | 99.7 | A&D FECG | | | | 5 |
| | 99.3 | NI-FECG challenge | | | | 69 |
| | 94.7 | Train on A&D FECG and test on NI-FECG challenge | | | | 69 |
| | 97.9 | Train on A&D FECG and test on NI-FECG | | | | 14 |

TS: Template subtraction, ES-RNN: echo state for the recurrent neural network, PCA: principal component analysis, LMS: Least Mean Square, RLS: Recursive Least Square, ICA: independent component analysis; OBACKC: optimized blind adaptive filtering using convolution kernel compensation; SVD-SW: singular value decomposition and smooth windowing.

which same to Warmerdam. Nevertheless, the better comparison needs more performance indices and analysis.

The effect of noise on the proposed method is evaluated by adding EMG noise with different amplitudes to the MECG signal to evaluate FECG generation. Figure 6 depicts the SNR of signals and the R-square acquired for subject 1 in the A&D FECG dataset based on training on other subjects with noisy signals.

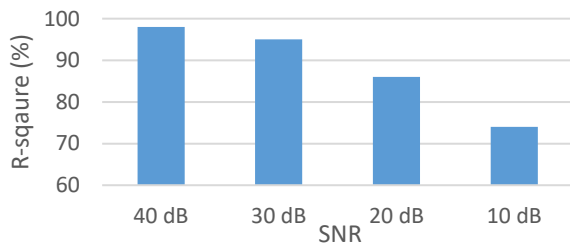

Fig. 6. The relation between SNR and R-square for subject 1 in the A&D FECG dataset.

It is essential to mention that QRS detection is not the primary task supported by the proposed method. The proposed method can extract the complete FECG signal, and a QRS detection algorithm is added on top of it to demonstrate one of the capabilities of the proposed method. The WEDD index, which shows distortion between predicted FECG and original FECG, can show that the proposed method can extract important waveforms (Table I and Figure 4). This capability can stimulate the idea that a similar attention mechanism can also be used for ECG wave analysis or to map other signals, such as ECG to PPG or vice versa [60].

One of the limitations of this study is the complexity of the model, common in deep learning algorithms. This complexity may increase the computational cost and maybe a burden for embedded systems, which can be eliminated by advances in hardware technology and using GPUs with less power consumption. The total training time is 10400 seconds, representing 208 seconds for each epoch. After training, the processing of 1000 samples requires a total of 0.14 seconds. Also, a larger sample size is required to show whether this approach can operate efficiently in all circumstances.

## V. Conclusion

A novel architecture based on the attention layer, sine activation function and cycle generative adversarial neural network is investigated to map maternal and fetal ECG. The proposed method is evaluated in two forms. First, the quality of FECG extracted from MECG is evaluated. Second, the fetal QRS detection from MECG is assessed. On the abdominal and direct FECG (A&D FECG) dataset, an average 98% R-Square [CI 95%: 97%, 99%] as the goodness of fit and 99.7 % F1-score [CI 95%: 97.8%, 99.9%] for QRS estimation achieved based on subject-leave-out validation. Besides A&D FECG dataset, the non-invasive FECG (NI-FECG) and NI-FECG challenge datasets are also used for fetal QRS estimation and achieved 99.7 % F1-score [CI 95%: 97.8-99.9], 99.6% F1-score [CI 95%: 98.2%, 99.9%] and 99.3% F1-score [CI 95%: 95.3%, 99.9%], respectively. A synthetic dataset is generated for investigating the effect of maternal and fetal heart rates on the performance which showed that the proposed method can be used in various fetal and maternal heart rate variations. Such results are comparable and superior to the-state-of-the-art results and therefore the proposed method is promising for FECG extraction.

## References

[1] P. R. Jeffries, S. Woolf, and B. Linde, "Technology-based vs. traditional instruction: A comparison of two methodsfor teaching the skill of performing a 12-lead ecg," *Nurs. Educ. Perspect.*, vol. 24, no. 2, pp. 70–74, 2003.

[2] M. A. Hasan, M. I. Ibrahimy, and M. B. I. Reaz, "Techniques of FECG signal analysis: detection and processing for fetal monitoring," *WIT Trans. Biomed. Health*, vol. 12, pp. 295–305, 2007.

[3] G. D. Clifford, F. Azuaje, and P. Mcsharry, "ECG statistics, noise, artifacts, and missing data," *Adv. Methods Tools ECG Data Anal.*, vol. 6, p. 18, 2006.

[4] E. R. Ferrara and B. Widraw, "Fetal electrocardiogram enhancement by time-sequenced adaptive filtering," *IEEE Trans. Biomed. Eng.*, no. 6, pp. 458–460, 1982.

[5] W. Zhong, L. Liao, X. Guo, and G. Wang, "A deep learning approach for fetal QRS complex detection," *Physiol. Meas.*, vol. 39, no. 4, p. 045004, 2018.

[6] M. Varanini, G. Tartarisco, R. Balocchi, A. Macerata, G. Pioggia, and L. Billeci, "A new method for QRS complex detection in multichannel ECG: Application to self-monitoring of fetal health," *Comput. Biol. Med.*, vol. 85, pp. 125–134, 2017.

[7] R. A. Shepoval'nikov, A. P. Nemirko, A. N. Kalinichenko, and V. V. Abramchenko, "Investigation of time, amplitude, and frequency parameters of a direct fetal ECG signal during labor and delivery," *Pattern Recognit. Image Anal.*, vol. 16, no. 1, pp. 74–76, 2006.

[8] G. D. Clifford, I. Silva, J. Behar, and G. B. Moody, "Non-invasive fetal ECG analysis," *Physiol. Meas.*, vol. 35, no. 8, p. 1521, 2014.

[9] B. Rafaely and S. J. Elliot, "A computationally efficient frequency-domain LMS algorithm with constraints on the adaptive filter," *IEEE Trans. Signal Process.*, vol. 48, no. 6, pp. 1649–1655, 2000.

[10] D. Mumford and A. Desolneux, *Pattern theory: the stochastic analysis of real-world signals*. CRC Press, 2010.

[11] M. Niknazar, B. Rivet, and C. Jutten, "Fetal ECG extraction by extended state Kalman filtering based on single-channel recordings," *IEEE Trans. Biomed. Eng.*, vol. 60, no. 5, pp. 1345–1352, 2013.

[12] J. Behar, A. Johnson, G. D. Clifford, and J. Oster, "A comparison of single channel fetal ECG extraction methods," *Ann. Biomed. Eng.*, vol. 42, no. 6, pp. 1340–1353, 2014.

[13] R. Martinek *et al.*, "Comparative effectiveness of ICA and PCA in extraction of fetal ECG from abdominal signals: Toward non-invasive fetal monitoring," *Front. Physiol.*, vol. 9, p. 648, 2018.

[14] M. R. Mohebbian, M. W. Alam, K. A. Wahid, and A. Dinh, "Single channel high noise level ECG deconvolution using optimized blind adaptive filtering and fixed-point convolution kernel compensation," *Biomed. Signal Process. Control*, vol. 57, p. 101673, 2020.

[15] N. Zhang *et al.*, "A novel technique for fetal ECG extraction using single-channel abdominal recording," *Sensors*, vol. 17, no. 3, p. 457, 2017.

[16] W. Zhong, X. Guo, and G. Wang, "QRStree: A prefix tree-based model to fetal QRS complexes detection," *PloS One*, vol. 14, no. 10, 2019.

[17] A. Josko and R. J. Rak, "Effective simulation of signals for testing ECG analyzer," *IEEE Trans. Instrum. Meas.*, vol. 54, no. 3, pp. 1019–1024, 2005.

[18] J. Behar, F. Andreotti, S. Zaunseder, Q. Li, J. Oster, and G. D. Clifford, "An ECG simulator for generating maternal-foetal activity mixtures on abdominal ECG recordings," *Physiol. Meas.*, vol. 35, no. 8, p. 1537, 2014.

[19] M. Frid-Adar, I. Diamant, E. Klang, M. Amitai, J. Goldberger, and H. Greenspan, "GAN-based synthetic medical image augmentation for increased CNN performance in liver lesion classification," *Neurocomputing*, vol. 321, pp. 321–331, 2018.

[20] L. Ying, A. Hertzmann, H. Biermann, and D. Zorin, "Texture and shape synthesis on surfaces," in *Rendering Techniques 2001*, Springer, 2001, pp. 301–312.

[21] Y. Aytar, L. Castrejon, C. Vondrick, H. Pirsiavash, and A. Torralba, "Cross-modal scene networks," *IEEE Trans. Pattern Anal. Mach. Intell.*, vol. 40, no. 10, pp. 2303–2314, 2017.

[22] M.-Y. Liu and O. Tuzel, "Coupled generative adversarial networks," in *Advances in neural information processing systems*, 2016, pp. 469–477.

[23] T. Kaneko, H. Kameoka, K. Tanaka, and N. Hojo, "Cyclegan-vc2: Improved cyclegan-based non-parallel voice conversion," in *ICASSP 2019-2019 IEEE International Conference on Acoustics, Speech and Signal Processing (ICASSP)*, 2019, pp. 6820–6824.

[24] S. Zhao *et al.*, "CycleEmotionGAN: Emotional semantic consistency preserved cycleGAN for adapting image emotions," in *Proceedings of the AAAI Conference on Artificial Intelligence*, 2019, vol. 33, pp. 2620–2627.

[25] O. Tmenova, R. Martin, and L. Duong, "CycleGAN for style transfer in X-ray angiography," *Int. J. Comput. Assist. Radiol. Surg.*, vol. 14, no. 10, pp. 1785–1794, 2019.

[26] B. Hou, J. Yang, P. Wang, and R. Yan, "LSTM-Based Auto-Encoder Model for ECG Arrhythmias Classification," *IEEE Trans. Instrum. Meas.*, vol. 69, no. 4, pp. 1232–1240, 2019.

[27] B. Taji, A. D. Chan, and S. Shirmohammadi, "False alarm reduction in atrial fibrillation detection using deep belief networks," *IEEE Trans. Instrum. Meas.*, vol. 67, no. 5, pp. 1124–1131, 2017.

[28] T. H. Linh, S. Osowski, and M. Stodolski, "On-line heart beat recognition using Hermite polynomials and neuro-fuzzy network," *IEEE Trans. Instrum. Meas.*, vol. 52, no. 4, pp. 1224–1231, 2003.

[29] J. Jezewski, A. Matonia, T. Kupka, D. Roj, and R. Czabanski, "Determination of fetal heart rate from abdominal signals: evaluation of beat-to-beat accuracy in relation to the direct fetal electrocardiogram," *Biomed. Tech. Eng.*, vol. 57, no. 5, pp. 383–394, 2012.

[30] R. Nurani, E. Chandraharan, V. Lowe, A. Ugwumadu, and S. Arulkumaran, "Misidentification of maternal heart rate as fetal on cardiotocography during the second stage of labor: the role of the fetal electrocardiograph," *Acta Obstet. Gynecol. Scand.*, vol. 91, no. 12, pp. 1428–1432, 2012.

[31] A. L. Goldberger *et al.*, "Components of a new research resource for complex physiologic signals," *PhysioBank PhysioToolkit Physionet*, 2000.

[32] I. Silva *et al.*, "Noninvasive fetal ECG: the PhysioNet/computing in cardiology challenge 2013," in *Computing in Cardiology 2013*, 2013, pp. 149–152.

[33] M. Varanini, G. Tartarisco, L. Billeci, A. Macerata, G. Pioggia, and R. Balocchi, "An efficient unsupervised fetal QRS complex detection from abdominal maternal ECG," *Physiol. Meas.*, vol. 35, no. 8, p. 1607, 2014.

[34] J. Behar, J. Oster, and G. D. Clifford, "Combining and benchmarking methods of foetal ECG extraction without maternal or scalp electrode data," *Physiol. Meas.*, vol. 35, no. 8, p. 1569, 2014.

[35] P. E. McSharry, G. D. Clifford, L. Tarassenko, and L. A. Smith, "A dynamical model for generating synthetic electrocardiogram signals," *IEEE Trans. Biomed. Eng.*, vol. 50, no. 3, pp. 289–294, 2003.

[36] E. C. G. An, "model for simulating maternal-foetal activity mixtures on abdominal ECG recordings/Behar J., Andreotti F., Zaunseder S. et al," *Physiol Meas*, vol. 35, no. 8, pp. 1537–1550, 2014.

[37] S. P. Von Steinburg *et al.*, "What is the 'normal' fetal heart rate?," *PeerJ*, vol. 1, p. e82, 2013.

[38] J. Patrick, K. Campbell, L. Carmichael, R. Natale, and B. Richardson, "Daily relationships between fetal and maternal heart rates at 38 to 40 weeks of pregnancy.," *Can. Med. Assoc. J.*, vol. 124, no. 9, p. 1177, 1981.

[39] J. J. Bailey *et al.*, "Recommendations for standardization and specifications in automated electrocardiography: bandwidth and digital signal processing. A report for health professionals by an ad hoc writing group of the Committee on Electrocardiography and Cardiac Electrophysiology of the Council on Clinical Cardiology, American Heart Association.," *Circulation*, vol. 81, no. 2, pp. 730–739, 1990.

[40] R. Sameni and G. D. Clifford, "A review of fetal ECG signal processing; issues and promising directions," *Open Pacing Electrophysiol. Ther. J.*, vol. 3, p. 4, 2010.

[41] S. Das and M. Chakraborty, "QRS detection algorithm using Savitzky-Golay filter," *ACEEE Int J Signal Image Process.*, vol. 3, no. 01, pp. 55–58, 2012.

[42] N. Rastogi and R. Mehra, "Analysis of Savitzky-Golay filter for baseline wander cancellation in ECG using wavelets," *Int J Eng Sci Emerg Technol*, vol. 6, no. 1, pp. 15–23, 2013.

[43] S. Patro and K. K. Sahu, "Normalization: A preprocessing stage," *ArXiv Prepr. ArXiv150306462*, 2015.

[44] A. Vaswani *et al.*, "Attention is all you need," *ArXiv Prepr. ArXiv170603762*, 2017.

[45] V. Sitzmann, J. Martel, A. Bergman, D. Lindell, and G. Wetzstein, "Implicit neural representations with periodic activation functions," *Adv. Neural Inf. Process. Syst.*, vol. 33, 2020.

[46] X. Fang, M. Wang, A. Shamir, and S.-M. Hu, "Learning Explicit Smoothing Kernels for Joint Image Filtering," in *Computer Graphics Forum*, 2019, vol. 38, no. 7, pp. 181–190.

[47] D. P. Kingma and J. Ba, "Adam: A method for stochastic optimization," *ArXiv Prepr. ArXiv14126980*, 2014.

[48] P. M. Bossuyt *et al.*, "STARD 2015: an updated list of essential items for reporting diagnostic accuracy studies," *Clin. Chem.*, vol. 61, no. 12, pp. 1446–1452, 2015.

[49] G. S. Collins, J. B. Reitsma, D. G. Altman, and K. G. Moons, "Transparent Reporting of a Multivariable Prediction Model for Individual Prognosis or Diagnosis (TRIPOD) The TRIPOD Statement," *Circulation*, vol. 131, no. 2, pp. 211–219, 2015.

[50] J. Miles, "R squared, adjusted R squared," *Wiley StatsRef Stat. Ref. Online*, 2014.

[51] S. Mehta *et al.*, "Performance of intraclass correlation coefficient (ICC) as a reliability index under various distributions in scale reliability studies," *Stat. Med.*, vol. 37, no. 18, pp. 2734–2752, 2018.

[52] J. M. Bland and D. G. Altman, "Statistical methods for assessing agreement between two methods of clinical measurement," *Int. J. Nurs. Stud.*, vol. 47, no. 8, pp. 931–936, Aug. 2010, doi: 10.1016/j.ijnurstu.2009.10.001.

[53] M. S. Manikandan and S. Dandapat, "Wavelet energy based diagnostic distortion measure for ECG," *Biomed. Signal Process. Control*, vol. 2, no. 2, pp. 80–96, 2007.

[54] J. Pan and W. J. Tompkins, "A real-time QRS detection algorithm," *IEEE Trans. Biomed. Eng.*, no. 3, pp. 230–236, 1985.

[55] J. F. Guerrero-Martinez, M. Martinez-Sober, M. Bataller-Mompean, and J. R. Magdalena-Benedito, "New algorithm for fetal QRS detection in surface abdominal records," in *2006 Computers in Cardiology*, 2006, pp. 441–444.

[56] S. Ghosh and Y. Rudy, "Application of l1-norm regularization to epicardial potential solution of the inverse electrocardiography problem," *Ann. Biomed. Eng.*, vol. 37, no. 5, pp. 902–912, 2009.

[57] X. Xu, J. Li, Y. Yang, and F. Shen, "Towards Effective Intrusion Detection Using Log-cosh Conditional Variational AutoEncoder," *IEEE Internet Things J.*, 2020.

[58] W. Zhong, L. Liao, X. Guo, and G. Wang, "Fetal electrocardiography extraction with residual convolutional encoder–decoder networks," *Australas. Phys. Eng. Sci. Med.*, vol. 42, no. 4, pp. 1081–1089, 2019.

[59] G. J. Warmerdam, R. Vullings, L. Schmitt, J. O. Van Laar, and J. W. Bergmans, "Hierarchical probabilistic framework for fetal R-peak detection, using ECG waveform and heart rate information," *IEEE Trans. Signal Process.*, vol. 66, no. 16, pp. 4388–4397, 2018.

[60] P. Sarkar and A. Etemad, "CardioGAN: Attentive Generative Adversarial Network with Dual Discriminators for Synthesis of ECG from PPG," *ArXiv Prepr. ArXiv201000104*, 2020.